\documentclass[letter]{aa} % for the letters 
\usepackage{graphicx}
\usepackage{txfonts}
\usepackage{hyperref}
\usepackage{graphicx}	% Including figure files
\usepackage{amsmath}	% Advanced maths commands
\usepackage{amssymb}	% Extra maths symbols
\usepackage{natbib}
\usepackage{longtable}
\usepackage{float}
\usepackage{threeparttable}
\include{def_bibtex}
\begin{document} 

   \title{Discovery of accretion-driven pulsations in the prolonged low X-ray luminosity state of the Be/X-ray transient GX~304-1}

   \author{A.~Rouco Escorial
          \inst{1}\fnmsep\thanks{A.RoucoEscorial@uva.nl}
          \and
          J.~van den Eijnden\inst{1}
          \and
          R.~Wijnands\inst{1}
          }

   \institute{Anton Pannekoek Institute for Astronomy, University of Amsterdam, Science Park 904, 1098 XH, Amsterdam, The Netherlands}

   \date{Received; accepted}

 \abstract{We present our \textit{Swift} monitoring campaign of the slowly rotating neutron star Be/X-ray transient GX~304-1 (spin period of $\sim$275\,s) when the source was not in outburst. We found that between its type-I outbursts the source recurrently exhibits a slowly decaying low-luminosity state (with luminosities of $10^{34-35}$\,erg~s$^{-1}$). This behaviour is very similar to what has been observed for another slowly rotating system, GRO~J1008-57. For that source, this low-luminosity state has been explained in terms of accretion from a non-ionised (`cold') accretion disk. Due to the many similarities between both systems, we suggest that GX~304-1 enters a similar accretion regime between its outbursts. The  outburst activity of GX~304-1 ceased in 2016. Our continued monitoring campaign shows that the source is in a quasi-stable low-luminosity state (with luminosities a few factors lower than previously seen) for at least one year now. Using our {\it NuSTAR} observation in this state, we found pulsations at the spin period, demonstrating that the X-ray emission is due to accretion of matter onto the neutron star surface. If the accretion geometry during this quasi-stable state is the same as during the cold-disk state, then matter indeed reaches the surface (as predicted) during this latter state. We discuss our results in the context of the cold-disk accretion model.}


\titlerunning{Pulsations in the low X-ray luminosity state of GX~304-1}
\authorrunning{Rouco Escorial et al.}
   \maketitle

\section{Introduction}\label{sec:GX304_introduction}

In high-mass X-ray binaries, a compact object is accreting from a massive companion (with a mass of $>$10\,$M_\odot$). The most common systems are the neutron star (NS) Be/X-ray transients in which magnetised NSs (with surface magnetic-field strengths of B$\sim$10$^{12-13}$~G) accrete matter from the decretion disks of their Be-type companions during outbursts (for a review see \citealt{Reig2011}). These systems can exhibit `normal', type-I outbursts and, in addition, sometimes `giant',  type-II outbursts. The type-I outbursts occur at periastron passages when the NS moves through the decretion disk of the Be star and accretes matter. In these outbursts, sources exhibit X-ray luminosities (L$_\textnormal{X}$) of $\sim$10$^{36-37}$\,erg~s$^{-1}$. The type-II outbursts are brighter than the normal ones and can reach luminosities of $\sim$10$^{38-39}$\,erg~s$^{-1}$. Their duration is also longer than type-I outbursts, generally lasting more than an orbital period. The nature of the mechanism(s) behind these type-II outbursts is unclear (for possible models see  \citealt{Moritani2013,Martin2014,Monageng2017,Laplace2017}).

The high luminosities exhibited by Be/X-ray transients during outbursts allow for detailed studies of their behaviour. Consequently, their outburst phenomenology is well known.  When not in outburst, their luminosities are significantly lower, and studying their behaviours becomes more difficult.  However, it is clear that the NS spin and magnetic field  play important roles in the phenomenology that these systems display at low luminosities. It is expected that below a certain accretion rate, matter cannot reach the NS surface anymore. This is due to the pressure exerted by the rotating NS magnetic field that expels the matter away; the systems then enter the so-called propeller regime (\citealt{Illarionov1975}; \citealt{Romanova2004}; \citealt{DAngelo2010}). After that, such systems may only exhibit very low luminosities ($< 10^{33}$\,erg~s$^{-1}$). 

However, in the case of slow rotating Be/X-ray transients (with typical spin periods P$_\textnormal{spin}$ of several tens of seconds and magnetic field strengths of $10^{12-13}$ G), it has been observed that several sources show an intermediate bright state (L$_\textnormal{X}$\,$\sim$10$^{34-35}$\,erg~s$^{-1}$) between their outbursts (e.g., \citealt{Tsygankov2017a}; \citealt{Ducci2018}; \citealt{Reig2018}). \citet{Tsygankov2017a} introduced a scenario to explain the observed X-ray emission during this state for the slowly rotating Be/X-ray transient GRO~J1008-57 (P$_\textnormal{spin}$\,$\sim$93.6\,s; \citealt{Stollberg1993}). For such slowly rotating systems, below a certain accretion rate and before the matter of the accretion disk is ejected by the propeller effect, the temperature of the matter in this disk may drop below the ionisation temperature of hydrogen. This results in a disk with a low degree of ionisation (called a `cold disk'), which can penetrate the magnetic field more easily than a hot ionised disk. The cold disk can move relatively close to the NS before it becomes hot again, causing the matter to be channelled by the  magnetic field to its poles. This might lead to observable  pulsations \citep{Tsygankov2017b}. So far, the long-term evolution of the cold-disk state has been poorly studied. In this letter, we present our X-ray monitoring campaign of the Be/X-ray transient GX~304-1 when it was likely accreting from a cold disk.

GX~304-1 is located at a distance of 2.01$^{+0.14}_{-0.13}$~kpc\footnote{The source is known with source Identifier~863533199843070208 in the Second \textit{Gaia} Data Release, GDR2 \citep{Gaia2018}. From this we estimated the distance following \citet{Bailer-Jones2018}.}. Its NS  has a spin period of $\sim$275\,s (\citealt{McClintock1977,Sugizaki2015})\footnote{See \href{https://gammaray.nsstc.nasa.gov/gbm/science/pulsars/lightcurves/gx304m1.html}{https://gammaray.nsstc.nasa.gov/gbm/science/pulsars/lightcurves/gx304m1.html} for the most recently observed spin period of the source (using the {\it FERMI} Gamma-ray Burst monitor). \label{foonoteFermi}} and a surface magnetic-field strength of 
 $\sim$4.7$\times$10$^{12}$\,G \citep{Yamamoto2011,Rothschild2017}. GX~304-1 is characterized by periods of strong activity wherein type-I outbursts recur periodically with a period of $\sim$132.2 days (the orbital period; \citealt{Sugizaki2015}), and periods wherein the source exhibited hardly any to no activity \citep{Priedhorsky1983,Pietsch1986,Sugizaki2015,Malacaria2017}. In particular, the source remained dormant since the early 1980's to 2008, when it showed renewed activity \citep{Manousakis2008}. The last reported outburst occurred in May 2016 \citep{Nakajima2016,Sguera2016,Rouco2016} and since then the source has remained in a low-luminosity state. (Fig.\,\ref{fig:GX304_combine}).
\section{Observations, analysis and results}\label{sec:GX304_observations_analysis_results}

We have monitored GX~304-1 using the X-ray Telescope (XRT; for $\sim$84.7\,ks) aboard the Neil Gehrels \textit{Swift} observatory (hereafter \textit{Swift}) to investigate its behaviour outside outbursts (ObsIDs 35072 and 88780). We have also intensively monitored the source since October 2017 when it became clear that its outburst activity had stopped. We also obtained a \textit{NuSTAR} observation (Section \ref{subsec:GX304_timing}) to study its spectrum above $>$10 keV and search for pulsations during its low-luminosity state.

\subsection{Light curves of GX~304-1}\label{subsec:GX304_lightcurve}

\begin{figure*}
	\includegraphics[width=2\columnwidth]{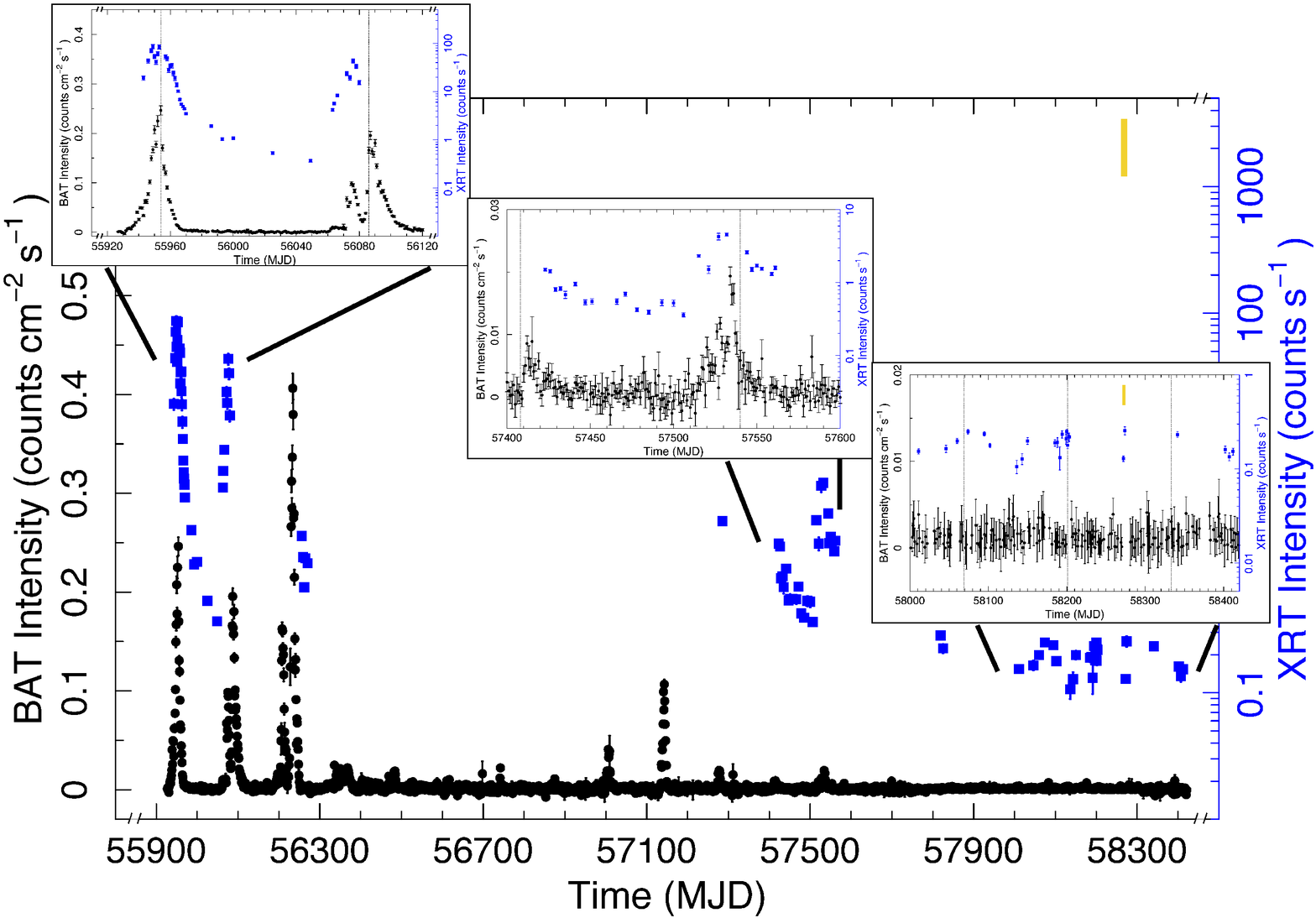}
    \caption{The BAT (15-50\,keV; black circles) and XRT (0.5-10\,keV; blue squares) light curves of GX~304-1 from January 1, 2012, to October 30, 2018. The time of our \textit{NuSTAR} observation is indicated with the yellow lines. The three insets show the behaviour of the source during different states (see main text for details). Vertical lines correspond to periastron passages (determined using the ephemeris of \citealt{Sugizaki2015}).}
    \label{fig:GX304_combine}
\end{figure*}

In Fig.\,\ref{fig:GX304_combine}, we show the light curves of the source obtained using the Burst Alert Telescope (BAT) aboard {\it Swift} (from the BAT transient monitor web page\footnote{\href{https://swift.gsfc.nasa.gov/results/transients/weak/GX304-1/}{https://swift.gsfc.nasa.gov/results/transients/weak/GX304-1/}}; \citealt{Krimm2013}) and the XRT (produced with the XRT products web interface\footnote{\href{http://www.swift.ac.uk/user\_objects/}{http://www.swift.ac.uk/user\_objects/}}; \citealt{Evans2009}). In the left inset, we show the first two outbursts  exhibited by the source in 2012. The maximum observed BAT count rates were $\sim$0.25 and $\sim$0.20\,counts~cm$^{-2}$~s$^{-1}$, respectively for the first and the second outbursts (i.e., the second peak of the second outburst). The XRT was used to monitor the evolution of both outbursts and the interval between them. The maximum observed XRT count rates were $\sim$87\,counts~s$^{-1}$ for the first outburst and $\sim$44\,counts~s$^{-1}$ for the second one (i.e., for the first peak of this outburst).  After the initial fast decay at the end of the first outburst, the source entered a state in which it decreased at a much slower rate: the XRT count rate dropped from $\sim$3.5 to $\sim$0.37\,counts~s$^{-1}$ in $\sim$79\,days (between MJD~55970 and 56049). No further count rate evolution could be investigated because the second outburst started. These count rates correspond to 0.5-100 keV luminosities of $\sim2.7 \times 10^{35}$ and $2.8 \times 10^{34}$ erg s$^{1}$, respectively. These luminosities were calculated using the spectral analysis and luminosities reported in \citet{Tsygankov2018}, when the source was even fainter, and then scaled using our observed XRT count rates (see below; this assumes that the spectral shape does not change at such low luminosities; this is consistent with what we can infer from our low quality XRT data).

Following the two outbursts reported above, a brighter one was detected  around MJD~56234 (Fig.\,\ref{fig:GX304_combine}; with peak BAT count rate of $\sim$0.41\,counts~cm$^{-2}$~s$^{-1}$). After this outburst, several additional outbursts were observed, but none of them were as bright as the 2012 ones (Fig.\,\ref{fig:GX304_combine}). No XRT data were obtained for several years during this period. However, on February 5, 2016, (MJD~57423) we obtained additional XRT observations after we noticed a small outburst in the BAT  on MJD~57415 (peaking at $\sim$0.89$\times$10$^{-2}$\,counts~cm$^{-2}$~s$^{-1}$; Fig.\,\ref{fig:GX304_combine}, middle inset). This outburst was followed by a brighter outburst between MJD~57510 and 57550. From the middle inset of Fig.\,\ref{fig:GX304_combine}, we can see that the source behaviour between these two outbursts is similar to what we observed in 2012. The observed XRT count rate decreased from $\sim$1.5 (on MJD~57423) to $\sim$0.36 counts~s$^{-1}$ (on MJD~57506) in $\sim$83 days (corresponding 0.5-100 keV luminosities of $\sim$11.5 and $\sim$2.8 $\times 10^{34}$ erg s$^{-1}$; using the method described earlier). This drop in count rate is lower than the one observed in-between the 2012 outbursts due to the lower count rate at the start of this phase. However, the end count rates are remarkably similar. We note that the overall trend during the 2016 low-luminosity state appears less smooth and with more variability, than what we observed in 2012.

After June 2016, GX~304-1 did not exhibit any detectable outbursts. When it was clear that the source indeed was not exhibiting outbursts anymore (after a few orbital cycles), we started an additional XRT monitoring campaign (started on MJD~58011; September 15, 2017) to investigate the overall behavior of the source (i.e, to determine if the source exhibited any increase in activity at periastron, and to determine if it would decay to fainter levels than previously observed). The results of our campaign are shown in the right inset in Fig.\,\ref{fig:GX304_combine}. We indeed observed the source at lower count rates than ever seen before, but we did not observe a clear overall trend in activity level (neither a decrease nor an increase). The source is quasi-stable (for over a year now) with count rates of $\sim$1-2.5$\times$10$^{-1}$\,counts~s$^{-1}$ with only a factor of 2-3 variability  (this corresponds to 0.5-100 keV luminosities of $\sim$0.8-1.9$\times10^{34}$\,erg~s$^{-1}$; determined using the method outlined above).  Although it appears that the count rate increased slightly during the several periastron passages that we monitored, similar count rate increases were also observed at other orbital phases (i.e., also at apoastron). Therefore, these fluctuations could just be random occurrences. Moreover, we planned our {\it NuSTAR} observation (Section \ref{subsec:GX304_timing}) at apoastron (to make sure the source was not in outburst) and we had several XRT observations scheduled close in time. The XRT count rate increased from $\sim$1.3$\times$10$^{-1}$\,counts~s$^{-1}$ (this XRT observation was simultaneously with our {\it NuSTAR} one) to 2.6$\times$10$^{-1}$\,counts~s$^{-1}$ within only a day. 

\subsection{Timing analysis of the \textit{NuSTAR} observation}\label{subsec:GX304_timing}

We observed GX~304-1 using \textit{NuSTAR} (\citealt{Harrison2013}) on June 3 (05:56:09 UTC), 2018, for $\sim$50\,ks with both FPMA and FPMB detectors (ObsID 90401326002). This observation was obtained to investigate the spectral behaviour of the source above 10~keV (reported in \citealt{Tsygankov2018}) and to search for pulsations. We ran the NUPIPELINE task (with SAAMODE=strict and TENTACLE=yes, due to the slightly high background event rates) to obtain clean event files and used the BARYCORR tool to perform the barycenter correction (using version 82 of the \textit{NuSTAR} clock correction). Finally, we obtained the light curves by means of NUPRODUCTS. In both observations, we used a circular region of 30 arcsec for extracting the source photons, and a 60 arcsec circular region for the background from a different chip (because of the background gradient that affected the chip where the source was located). Although we produced background-subtracted light curves, we restricted them to the 3–30\,keV energy range as the background dominates the source above that energy.

\begin{figure}
	\includegraphics[width=\columnwidth]{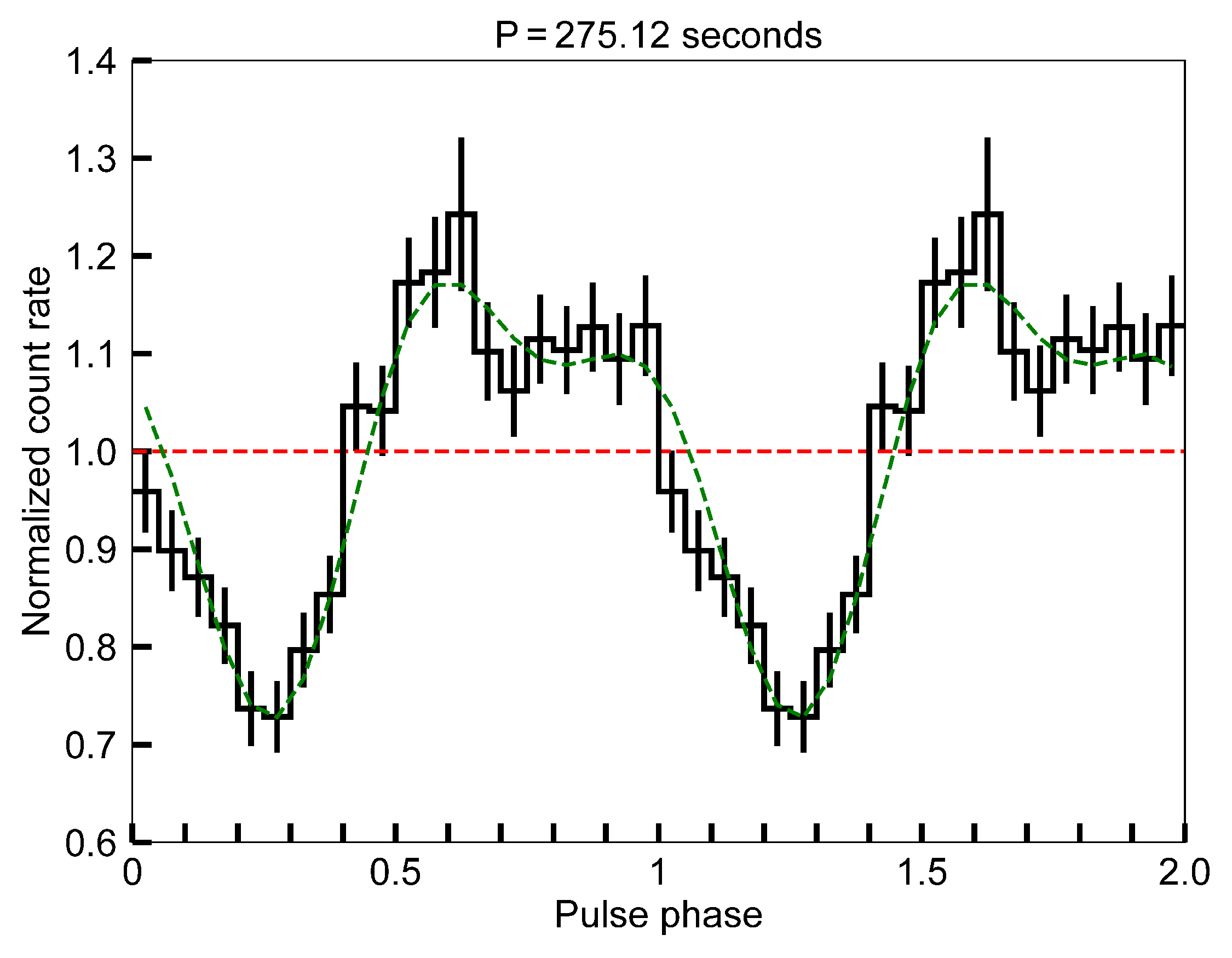}
    \caption{The \textit{NuSTAR} combined FPMA and FPMB light curve (3-30 keV)  folded on the best fitting period of \textbf{$P = 275.12$\,s.} The count rate has been normalized by dividing it by the mean count rate of $0.313$ count s$^{-1}$ (indicated by the red dashed line). The green dotted line shows the best fitting model consisting of a fundamental and one harmonic at twice the frequency.}
    \label{fig:GX304_folded_lc}
\end{figure}

We searched for pulsations in the {\it NuSTAR} observation using the phase folding method introduced in \citet{Leahy1983}, and used it as implemented in the \textsc{ftool}\footnote{\href{https://heasarc.gsfc.nasa.gov/ftools/ftools\_menu.html}{https://heasarc.gsfc.nasa.gov/ftools/ftools\_menu.html}} \textsc{efsearch}. Using a custom \textsc{python} script, we folded the light curve on a range of periods around the known period of $\sim$275\,s. The best fit period is defined as the period for which the $\chi^2$ of the folded light curve with respect to a constant is maximum. We clearly detected pulsations in the combined FPMA and FPMB light curve at P=275.12$\pm$0.02\,s (with 1$\sigma$ error). The pulsations are present in both individual detectors and are consistent with the known period of GX~304-1. The error on the period was determined following the approach in \citet{Brumback2018a, Brumback2018b}: using the best-fit period, we simulated $500$ fake sets of an FPMA and an FPMB light curve. For each set, we repeated our analysis and measured the period in the simulated data. We adopted the standard deviation of the obtained best-fit periods as the 1$\sigma$ error (as quoted above) on the measured period.

The light curve folded on the best-fitting period is shown in Fig.\,\ref{fig:GX304_folded_lc}. The profile can be described by a combination of a cosine function and a harmonic at half the period. A fit with such a model is shown in Fig.\,\ref{fig:GX304_folded_lc} as the green dotted line. This combined model implies a fundamental amplitude of $19.3\%$, while the harmonic contributes $8.6\%$. Taking a model-independent view, from Fig.\,\ref{fig:GX304_folded_lc} we conclude that the waveform varies over a range of $0.72$ to $1.25$ times the mean count rate.

\section{Discussion}\label{sec:GX304_discussion}

We have reported on {\it Swift}/XRT observations of the Be/X-ray transient GX~304-1, which harbours a slow X-ray pulsar ($\sim$275\,s), obtained when the source was in-between type-I outbursts in 2012, and after these outbursts ceased in 2016. Additionally, we report on the timing analysis of our \textit{NuSTAR} observation in the latter period. At all times, the source was clearly detected at luminosities of $\sim$10$^{34-35}$\,erg~s$^{-1}$, which are significantly lower than the ones observed during outbursts, but still relatively high compared to the much fainter luminosities observed in other Be/X-ray transients when not in outburst ($\sim10^{32-33}$ erg s$^{-1}$; \citealt{Tsygankov2017b} and references therein). However, such intermediate bright states have also been detected in several other, slowly spinning systems (e.g., \citealt{Haberl2016,Tsygankov2017a}). So far, only for one system, GRO~J1008-57, this state has been equally well-monitored \citep{Tsygankov2017a} as we did for GX~304-1. Remarkably, the behaviour is very similar in both sources: both exhibited luminosities in the range of $\sim10^{34-35}$\,erg~s$^{-1}$, which slowly declined between the adjacent type-I outbursts (e.g., clearly visible when comparing the left inset in Fig.\,\ref{fig:GX304_combine} with Fig.\,1a in \citealt{Tsygankov2017a}).  Therefore, it is quite likely that this behaviour is caused by the same physical mechanism in both sources. 

\citet{Tsygankov2017a} suggested that during this state the X-ray emission originates from accretion of matter down to the NS through a cold, non-ionised disk. This state can only occur for systems with slow spinning (with spin of several tens of seconds or slower) and magnetised ($\sim$10$^{12-13}$\,G) NSs, since only for such systems the accretion disk would become non-ionised before the propeller effect is initiated (see Section \ref{sec:GX304_introduction}). Since GX~304-1 spins slowly at $\sim$275\,s and its magnetic field is $\sim$4.7$\times$10$^{12}$\,G, we suggest that GX~304-1, similar to GRO~J1008-57, was accreting from such a disk between its type-I outbursts (as was proposed by \citealt{Tsygankov2017a}, see their Fig.~1a). 

However, alternative physical scenarios have been proposed to explain the low-luminosity behaviour in Be/X-ray transients: the two main ones are the cooling emission from an accretion-heated NS crust and residual low-level accretion onto the NS surface even when the system is in the propeller regime. In the NS crust cooling scenario, the crust would have been heated by the accretion of matter during the preceding outburst and in between outbursts this crust would cool resulting in observable emission. However, we observed GX~304-1 at luminosities of $\sim10^{34}$ erg s$^{-1}$ which are, at least, one order of magnitude higher than the ones observed in systems that might indeed have exhibited such crust cooling behaviour (L${_\textnormal{X}}$$\sim$10$^{32-33}$\,erg~s$^{-1}$; e.g. \citealt{Wijnands2016} and \citealt{Rouco2017}; see the review by \citealt{Wijnands2017}). In addition, we observed short term variability (on time scales of days) for GX~304-1 which is not expected in the cooling scenario. Therefore, we do not think that we observed NS crust cooling emission in GX~304-1. 

The other main alternative model to explain low-level emission in between outbursts is that in which the systems have entered the propeller regime but that still matter might reach the surfaces of the NSs (i.e., due to "leakage" of matter through the magnetosphere, although how exactly this would work is not fully understood; e.g., \citealt{Orlandini2004}; \citealt{Mukherjee2005}; \citealt{Rothschild2013}; \citealt{Doroshenko2014}). For this model to work, the source has to have entered the propeller regime but this would only happen for GX~304-1 at luminosities of $< 2 \times 10^{32}$ erg s$^{-1}$ (using Eq. 4 of \citealt{Tsygankov2017a} with a NS mass of 1.4 M$_\odot$ and a radius of 10 km) which is significantly lower than the actual luminosities we observed for this source. Therefore, we think that this is also not a viable model to explain our observed emission and we conclude that the cold disk hypothesis is the most compelling explanation for the low-luminosity state in GX~304-1.

In the case of GX~304-1, we have now also determined that this low-luminosity, cold-disk state is a recurrent phenomenon because we have now observed it in-between two sets of type-I outbursts (Fig.\,\ref{fig:GX304_combine}). During the outbursts, the accretion disk around the NS most likely fills up again to such a degree that, once the outbursts are over, the disk around the NS contains enough matter for the source to enter the cold-disk phase. In this respect, GX~304-1 is a very interesting source because, after June 2016, the system did not exhibit any outbursts anymore. One would expect that if the accretion disk around the NS is not being fed with matter in the absence of outbursts, the cold disk would slowly empty since all the matter would eventually be accreted onto the NS. Therefore, in this scenario, one would expect that the luminosity would slowly decrease during this phase (until all the matter in the cold disk is consumed and other emission mechanisms might take over; see also the discussion in \citealt{Tsygankov2017b}).

To test this hypothesis, we set up a XRT campaign to investigate the long-term behaviour of GX~304-1 after it was clear that the source did not exhibit any outbursts anymore. Indeed, we found that the luminosity had decreased by a factor of 2-3 compared to that observed in the cold-disk phase between type-I outbursts. However, we did not observe an overall decay trend as we expected. Instead, the source has now been in a quasi-stable state for over one year, in which its count rate only varies by a factor of 2-3. This variability does not seem to be correlated with periastron passages because similar variability is also observed at other orbital phases (i.e, at apoastron). The reasons for this quasi-stable state and the observed variability are unclear. It might be that during periastron passages the cold disk continues to be replenished, either because matter is transferred from the decretion disk or due to the wind of the Be star. In any case, it remains unclear why this extra matter does not cause a full type-I outburst or, at least, noticeable increases in luminosity at periastron. We continue our monitoring campaign to further study this enigmatic state in GX~304-1.

Although the luminosities observed for GX~304-1 and GRO~J1008-57, during the cold-disk state are of such high level that only accretion down to the inner regions could cause the emission, so far it was not proven that matter indeed reached their NSs. Accretion down to the surface would be demonstrated if pulsations were detected, and \citet{Tsygankov2017b} predicted that such phenomenon should be observed. Our detection of X-ray pulsations at the spin period of GX~304-1 during our \textit{NuSTAR} observation at the quasi-stable state of the source (performed at apoastron and, by coincidence, at one of the lowest observed luminosity), confirms the fact that, indeed, matter is still accreted all the way down to the surface. Since this quasi-stable state is, most likely, just an extension of the cold-disk phase, we now have strong evidence that matter is accreted down to the surface when these systems are in the cold-disk phase.

Our observed period is consistent with the one expected from the general spin-down trend that the source seems to follow, which started at the end of the strong type-I outburst activity in 2012 and continued to the present day (see the {\it Fermi}/GBM data as linked in footnote\,\ref{foonoteFermi}; see \citealt{Postnov2015} and \citealt{Sugizaki2015} for the spin evolution until 2013). The strength of the pulsations ($\sim20$\%) and the pulse profile are similar to the ones observed during outbursts \citep{Devasia2011,Malacaria2015,Jaisawal2016}. However, the pulse profiles vary strongly during and between outbursts, and with energy. Therefore, it is unclear whether the observed similarities between outburst and the quasi-stable state are due to the same underlying (inner) accretion geometry or, purely, due to chance. The fact that the source is currently in a quasi-stable state allows for additional observations to study the pulsations in more detail and to better understand accretion through a cold disk. In addition, pulsations are also expected in the cold-disk phase of other slowly rotating Be/X-ray transients (i.e., in GRO~J1008-57), therefore these systems are perfect targets to study this stage further. 

\begin{acknowledgements}\label{sec:GX304_acknowledgements}

ARE and RW acknowledge support from a NWO Top grant, module 1, awarded to RW. JvdE is supported by NWO. The authors thank Fiona Harrison and the \textit{NuSTAR} team for rapidly approving and executing our observation. We also thank the {\it Swift} team (and, i.e, Neil Gehrels and Brad Cenko) for granting and scheduling our (many) XRT observations.

\end{acknowledgements}

\bibliographystyle{aa}
\bibliography{references}\label{GX304_references}

\end{document}